\documentstyle{article}
\begin{document}
\title{On Pseudospherically Symmetric Repulsive \\
Gravitational Field}
\author{Luis A. Anchordoqui, Jos\'e  D. Edelstein\thanks{CONICET}, 
Carlos N\'u\~nez \\
{\it Departamento de F\'\i sica, Universidad Nacional de La Plata}\\
{\it C.C. 67, (1900) La Plata}\\
{\it Argentina}\\
and  \\
Graciela S. Birman$^{\ast}$ \\
{\it Departamento de Matem\'atica, Fac. Cs. Exactas, UNCPBA}\\
{\it Pinto 399, (7000) Tandil}\\ 
{\it Argentina}}
\date{}
\maketitle

\begin{abstract}
Solutions of Einstein vacuum equations, for a static
pseudospherically symmetric system, are presented. 
They describe a naked singularity and a singular solution with
many resemblances to the Schwartzschild solution but with two major
differences: its static region, lying inside the null horizon,
sees the singularity, and its effective gravitational field is  
repulsive. We shortly discuss on the phenomenological plausibility
of this last solution as a self-consistent system living on a
space-time domain, and discuss some features of particle geodesics
in its gravitational field. 
\end{abstract}
\vspace{1cm}

As a fundamental physical theory, General Relativity is well known
for not imposing any constraints on spacetime geometry
just but a Lorentzian-manifold structure and Einstein field
equations. In specific problems, it may be appropriate to impose
additional model-dependent constraints, such as a specific
symmetry or asymptotic conditions on the geometry.
Thus, the study of Einstein equations and their solutions have
attracted so much attention from the point of view of numerical 
analysis \cite{papa} as well as exact solutions coupled to matter
\cite{roy,ran}, solutions in diverse dimensions \cite{luis},
solitonic solutions \cite{bart}, etc.

In this contribution, we shall assume the existence of a region in
the Universe having spatial pseudospherical symmetry. This region
could be thought of as a kind of space-time domain possibly
originated in an anisotropy occured at the Universe expansion.
In any case, we shall solve under this hypothesis the Einstein vacuum
equations inside the domain, reaching fairly interesting results.
We obtain two possible singular solutions depending upon the sign of
an integration constant $\varpi$:

\noindent
(i)  $\varpi < 0$ leads to a naked singularity, while

\noindent
(ii) $\varpi > 0$ gives rise to a singular space-time with a 
null-horizon and a repulsive gravitational field. The static
region lies inside the horizon and, thus, sees the singularity.

We shall analyse these singular solutions, study particle
geodesics in the background metric of solution (ii), and
argue on the self-consistency of the resulting effective
repulsive gravitational field.

Let us start by considering the Einstein vacuum equations
in a space-time domain with spatial pseudospherical symmetry. 
The solution we shall develope, 
choosing a pseudospherical gauge, is based on the
following metric tensor,
\begin{equation}
g_{\mu \nu} = \; {\rm diag} \; ( e^{\nu}, \, -e^{\lambda}, \, -r^2, \; -r^2
e^{2 \theta}).  
\label{sol}
\end{equation}

In spite of the fact that $\nu$ and $\lambda$ are arbitrary functions of
$r$ and $t$, as in Schwarzschild's case 
\cite{Landau}, there is a
coordinate system where the metric tensor is static, in other words,
the universe is stationary for
a privileged observer that detects no change in the intrisic geometry
of the space. Concerning $\theta$, it is a pseudospherical parameter
valued in the interval $(-\infty,0]$ \cite{pseudo}. 
In such a coordinate system, 
the solution of Einstein vacuum equations reads
\begin{equation}
ds^2 = - \left( 1 - \frac{\varpi}{r} \right) dt^2 + \left( \frac{1}{1 -
\frac{\varpi}{r}} \right) dr^2 - r^2 ( d\theta^2 + e^{2 \theta}
d\phi^2 ).
\label{ds}
\end{equation}
A naked singularity rises up whenever the integration constant $\varpi$ 
is negative. 
We shall consider, in the following, $\varpi$ as a positive 
constant to be determined from continuity conditions on the boundary of 
the domain\footnote{Note that as our solution is valid on a bounded
region of spacetime, there is no ``Newtonian'' restriction on the
sign of $\varpi$.}.
In this case, it is easily seen from equation (\ref{ds}) that
there are pathologies at $r = \varpi$ and $r=0$. 
In order to clarify about the physical significance of
these divergences, let us compute the non-vanishing
components of the Riemann tensor. They are simply 
\begin{equation} 
R_{tr}\;^{tr} = R_{\theta\phi}\;^{\theta\phi}  = \frac{\varpi}{r^3} 
\;\;\;\;\; \mbox{and} \;\;\;\;\;
R_{t\theta}\;^{t\theta}  = R_{t\phi}\;^{t\phi} = 
R _{r\theta}\; ^{r\theta}= R_{r\phi}\;^{r\phi} = 
- \frac{\varpi}{2 \; r^3},
\label{riemann}
\end{equation}
showing that an imaginary test observer aproaching from infinite
must encounter infinite tidal forces at the origin, independently of 
the route he uses to reach there.
This is obvious from the curvature invariant $I$
\begin{equation}
I \equiv R_{\lambda\eta}\;^{\mu\nu}  R_{\mu\nu}\;^{\lambda\eta} = 12
\; \frac{\varpi^2}{r^6}.
\label{curvinv}
\end{equation}
On the other hand, the Riemann tensor is completely regular at
$r = \varpi$. Moreover, the proper distance from $r=\varpi$ to an 
arbitrary point $r$, is well defined in the interval $0<r<\infty$ 
as is shown in the following equations
\begin{equation}
\int^r_{\varpi} |g_{rr}|^{1/2} dr = [r (r - \varpi)]^{1/2} + 
\varpi \; \ln \left| \left( \frac{r}{\varpi} - 1 \right)^{1/2}
+  \left( \frac{r}{\varpi} \right)^{1/2} \right| 
\label{proper1}
\end{equation}
when  $r < \varpi $ whereas
\begin{equation}
\int^r_{\varpi} |g_{rr}|^{1/2} dr = - \varpi \; \; {\rm arccot} 
\left[ \frac{r^{1/2}}{(\varpi - r)^{1/2}} \right]
- [ r (\varpi - r )]^{1/2}
\label{proper2}
\end{equation}
when $ r > \varpi$.
It is worthwhile to point out that at $r = \varpi$ we have a 
reversal in the roles of $t$ and $r$ as timelike and spacelike 
coordinates.
The spacetime breaks into two parts, resembling the standard,
positive mass, Schwartzschild solution: a time-like static region with the
Killing vector in the time direction and a dynamic
region with three space-like symmetries; the boundary between these
regions, at $r = \varpi$, being a null-horizon. One major difference
with the Schwartzschild solution is that the static region of the
pseudospherical singularity is in the neighbourhood of the origin
and, therefore, sees the singularity.

In order to get a deeper insight into the pseudospherical solution,
it will be interesting to study the orbits followed by particles 
inmersed in its gravitational field (\ref{ds}). The particle
geodesics in this background metric will reveal several interesting
features of the pseudospherical singularity. We shall choose the 
coordinate system in such a way that the radial projection
of the orbits coincide with $\theta = 0$. In this oriented
coordinate system, the particle will have zero momentum in the 
$\theta$-direction.
The invariances \mbox{$t \rightarrow t + \Delta t$} and \mbox{$\phi
\rightarrow \phi + \Delta \phi$} define two Killing vectors which we
shall identify with the conjugate momenta $p_0 \equiv E$ (particle's
energy) and $p_{\phi} \equiv L$ (an angular momentum-like vector).
We will use these conserved quantities, as usual, to obtain a first
integral of the equations of motion.
Consistency with the equivalence principle, demands test particles to
follow the same world line regardless of its mass. Hence the
quantities relevant for the motion of particles are: the energy per
unit rest mass $\tilde{E} = E/m$ and the ratio $\tilde{L} \equiv L/m$.
Taking this into account, the on-shell condition of the particle 
can be written as
\begin{equation}
- \frac{\tilde{E}^2}{1 - \frac{\varpi}{r}} + \frac{1}{1 - \frac{\varpi}{r}}
\left( \frac{dr}{d\tau} \right) ^2 - \frac{\tilde{L}^2}{r^2} - 1 = 0,
\label{mass1}
\end{equation}
or, equivalently,
\begin{equation}
\left( \frac{dr}{d\tau} \right) ^2 = \tilde{E}^2 - \tilde{V}_{\rm eff}^2,
\label{mass2}
\end{equation}
with the effective potential given by
\begin{equation}
\tilde{V}_{\rm eff}^2 = \left( \frac{\varpi}{r} - 1 \right)
\left(\frac{\tilde{L}^2}{ r^2} + 1 \right).
\label{mass3}
\end{equation}
Interestingly enough, we see that the effective potential is repulsive 
for $\tilde{L} = 0$ while, contrary to the usual cases, the ``centrifugal'' 
term of $\tilde{V}^2_{\rm eff}$ gives rise to an attractive contribution to 
the effective potential. We will come back to this striking point later.

The turning point $R$ can be obtained by setting
$dr/d\tau=0$.
For a given energy $\tilde{E}_{\circ}$, 
restricting ourselves to radial motion ($\tilde{L}
\equiv 0 )$ this results in
\begin{equation}
R = \frac{\varpi}{\tilde{E}_{\circ}^2 + 1 }.
\label{turning}
\end{equation}
Then, the proper time can be simply expressed, from eq.(\ref{mass2}),
as
\begin{equation}
\tau = \int \frac{dr}{\varpi^{1/2}} \left( \frac{1}{R} - \frac{1}{r} 
\right)^{-1/2}.
\label{proper3}
\end{equation}
This integral can be evaluated using the suitable change of variables 
\begin{equation}
r = \frac{R}{2} ( 1 + \cosh \eta ),
\label{change}
\end{equation}
in such a way that the `proper time' it takes a particle to travel from 
$r = R$ to $r = 4\varpi$\footnote{We consider $r = 4 \varpi$ a radius 
where the particle is far away from the influence of the potential.} 
is a finite quantity given by
\begin{equation}
\tau = \frac{R}{2} \; \sqrt{\frac{R}{\varpi}} \; \left\{\;
{\rm arccosh} \left( \frac{8 \varpi}{R} - 1 \right) + 4 \;
\sqrt{\frac{ \varpi}{R} \left( \frac{ 4 \varpi}{R} - 1 \right)} \; 
\right\}.
\label{proper4}
\end{equation}
In fact, this quantity should be referred to as a proper distance,
being calculated inside the dynamical region.
In order to compute the `coordinate time' taken by the particle to 
reach the null-horizon, it will be useful to introduce appropriate
``tortoise coordinates'' \cite{Grav}
\begin{equation}
r^{\star} = - [r + \varpi \ln (\varpi - r)] \;\;\;\;\; \mbox{such that}
\;\;\;\;\; \frac{dr}{d\tau} \equiv \tilde{E} \frac{d r^{\star}}{dt}.
\label{tortoise1}
\end{equation}
It is easy to see from (\ref{mass3}) that $\tilde{E}^2 - \tilde{V}^2_{\rm 
eff} \cong \tilde{E}^2$ near $r = \varpi$. Then, after (\ref{mass2}) and
(\ref{tortoise1}), $dr^{\star}/dt \cong 1$. Thus, 
\begin{equation}
r^{\star} = - \{ r + \varpi \ln ( \varpi - r ) \} \cong \varpi \ln
(\varpi - r),
\label{tortoise2}
\end{equation}
or, equivalently,
\begin{equation}
r = \varpi - C e^{-t/\varpi}, 
\label{tortoise3}
\end{equation}
with $C \equiv$ constant. It can be seen from equations (\ref{mass3}) 
and (\ref{tortoise1}) that the approach of $\tilde{V}^2_{\rm eff}$ to 
zero as $r \rightarrow \varpi$ is seen to be exponential in
$r^{\star}$. In the same vein, equation (\ref{tortoise3}) shows that it 
takes infinite `coordinate time' to arrive at $r = \varpi$.

Let us briefly discuss, at this point, about the repulsive nature
of the pseudospherical solution. It has certainly appeared from the
lack of a `Newtonian' restriction on the bounded region of space-time
given by the spatial pseudospherically symmetric domain. This domain 
can be thought of as a kind of structure that might have appeared in 
the early universe generating a disconnected region, still unobserved, 
with pseudospherical geometry. It is important to stress that this 
symmetry could be generated by a mass distribution external to the 
domain. This fact is not forbidden, in principle, by Birkhoff's 
theorem which does not apply whenever symmetries are
not spherical. The boundary  of the domain could be inside the horizon, 
this leading to the appearence of a naked singularity. 

The main challenge posed by this singular pseudospherically symmetric
solution should be to find a matter system that would be able to 
generate this odd gravitational field, and to carefully study how it
smoothly glues with regular solutions outside the domain. This is,
of course, a hard problem that should be fulfilled in order to give
complete consistency to our pseudospherical solution into a
`Newtonian'-like universe. The study of this issue, as well as
a complete analysis of particle geodesics in the background metric
of the pseudospherical singularity, is being completed \cite{nous}.
We hope to be able to give an answer to the matching-conditions 
problem in the boundary of the domain in a forthcoming work.
\vspace{3mm}

We would like to thank H\'ector Vucetich for a critical 
reading of the manuscript.

\end{document}